\documentclass[twocolumn,pre]{revtex4}

\usepackage{dcolumn}
\usepackage{amsmath}

\usepackage{graphicx}

\def\d{{\rm d}}
\def\e{{\rm e}}
\def\i{{\rm i}}
\def\eref#1{(\protect\ref{#1})}
\def\etal{{\it{}et~al.}}
\def\Li{\mathop{\rm Li}}
\def\av#1{\left\langle#1\right\rangle}
\def\set#1{\left\lbrace#1\right\rbrace}

\newlength{\figurewidth}
\begin{document}

\title{The spread of epidemic disease on networks}
\author{M. E. J. Newman}
\affiliation{Center for the Study of Complex Systems,
University of Michigan, Ann Arbor, MI 48109--1120}
\affiliation{Santa Fe Institute, 1399 Hyde Park Road, Santa Fe, NM 87501}
\begin{abstract}
  The study of social networks, and in particular the spread of disease on
  networks, has attracted considerable recent attention in the physics
  community.  In this paper, we show that a large class of standard
  epidemiological models, the so-called susceptible/infective/removed (SIR)
  models can be solved exactly on a wide variety of networks.  In addition
  to the standard but unrealistic case of fixed infectiveness time and
  fixed and uncorrelated probability of transmission between all pairs of
  individuals, we solve cases in which times and probabilities are
  non-uniform and correlated.  We also consider one simple case of an
  epidemic in a structured population, that of a sexually transmitted
  disease in a population divided into men and women.  We confirm the
  correctness of our exact solutions with numerical simulations of SIR
  epidemics on networks.
\end{abstract}
\maketitle

\section{Introduction}
Many diseases spread through human populations by contact between infective
individuals (those carrying the disease) and susceptible individuals (those
who do not have the disease yet, but can catch it).  The pattern of these
disease-causing contacts forms a network.  In this paper we investigate the
effect of network topology on the rate and pattern of disease spread.

Most mathematical studies of disease propagation make the assumption that
populations are ``fully mixed,'' meaning that an infective individual is
equally likely to spread the disease to any other member of the population
or subpopulation to which they belong~\cite{Bailey75,AM91,Hethcote00}.  In
the limit of large population size this assumption allows one to write down
nonlinear differential equations governing, for example, numbers of
infective individuals as a function of time, from which solutions for
quantities of interest can be derived, such as typical sizes of outbreaks
and whether or not epidemics occur.  (Epidemics are defined as outbreaks
that affect a non-zero fraction of the population in the limit of large
system size.)  Epidemic behavior usually shows a phase transition with the
parameters of the model---a sudden transition from a regime without
epidemics to one with.  This transition happens as the ``reproductive
ratio'' $R_0$ of the disease, which is the fractional increase per unit
time in the number of infective individuals, passes though one.

Within the class of fully mixed models much elaboration is possible,
particularly concerning the effects of age structure in the population, and
population turnover.  The crucial element however that all such models lack
is network topology.  It is obvious that a given infective individual does
not have equal probability of infecting all others; in the real world each
individual only has contact with a small fraction of the total population,
although the number of contacts that people have can vary greatly from one
person to another.  The fully mixed approximation is made primarily in
order to allow the modeler to write down differential equations.  For most
diseases it is not an accurate representation of real contact patterns.

In recent years a large body of research, particularly within the
statistical physics community, has addressed the topological properties of
networks of various kinds, from both theoretical and empirical points of
view, and studied the effects of topology on processes taking place on
those networks~\cite{Strogatz01,AB02}.  Social
networks~\cite{WS98,ASBS00,Liljeros01,GN02}, technological
networks~\cite{ABW98,AJB99,FFF99,Broder00}, and biological
networks~\cite{Jeong00,Jeong01,FW00,WM00,MS02} have all been examined and
modeled in some detail.  Building on insights gained from this work, a
number of authors have pursued a mathematical theory of the spread of
disease on networks~\cite{MN00a,KA01,PV01a,MPV02,WSS01,Warren02}.  This is
also the topic of the present paper, in which we show that a large class of
standard epidemiological models can be solved exactly on networks using
ideas drawn from percolation theory.

The outline of the paper is as follows.  In Section~\ref{perc} we introduce
the models studied.  In Section~\ref{simplest} we show how percolation
ideas and generating function methods can be used to provide exact
solutions of these models on simple networks with uncorrelated transmission
probabilities.  In Section~\ref{correlated} we extend these solutions to
cases in which probabilities of transmission are correlated, and in
Section~\ref{structured} to networks representing some types of structured
populations.  In Section~\ref{concs} we give our conclusions.

\section{Epidemic models and percolation}
\label{perc}
The mostly widely studied class of epidemic models, and the one on which we
focus in this paper, is the class of susceptible/infective/removed or SIR
models.  The original and simplest SIR model, first formulated (though
never published) by Lowell Reed and Wade Hampton Frost in the 1920s, is as
follows.  A population of $N$ individuals is divided into three states:
susceptible~(S), infective~(I), and removed~(R).  In this context
``removed'' means individuals who are either recovered from the disease and
immune to further infection, or dead.  (Some researchers consider the R to
stand for ``recovered'' or ``refractory.''  Either way, the meaning is the
same.)  Infective individuals have contacts with randomly chosen
individuals of all states at an average rate $\beta$ per unit time, and
recover and acquire immunity (or die) at an average rate $\gamma$ per unit
time.  If those with whom infective individuals have contact are themselves
in the susceptible state, then they become infected.  In the limit of large
$N$ this model is governed by the coupled nonlinear differential
equations~\cite{Bailey75}:
\begin{equation}
{\d s\over\d t} = -\beta i s,\quad
{\d i\over\d t} = \beta i s - \gamma i,\quad
{\d r\over\d t} = \gamma i,
\label{sir}
\end{equation}
where $s(t)$, $i(t)$, and $r(t)$ are the fractions of the population in
each of the three states, and the last equation is redundant, since
necessarily $s+i+r=1$ at all times.  This model is appropriate for a
rapidly spreading disease that confers immunity on its survivors, such as
influenza.  In this article we will consider only diseases of this type.
Diseases that are endemic because they propagate on timescales comparable
to or slower than the rate of turnover of the population, or because they
confer only temporary immunity, are not well represented by this model;
other models have been developed for these cases~\cite{Hethcote00}.

The model described above assumes that the population is fully mixed,
meaning that the individuals with whom a susceptible individual has contact
are chosen at random from the whole population.  It also assumes that all
individuals have approximately the same number of contacts in the same
time, and that all contacts transmit the disease with the same probability.
In real life none of these assumptions is correct, and they are all grossly
inaccurate in at least some cases.  In the work presented here we remove
these assumptions by a series of modifications of the model.

First, as many others have done, we replace the ``fully mixed'' aspect with
a network of connections between
individuals~\cite{MN00a,KA01,PV01a,WSS01,Warren02,MPV02,SS88,Longini88,KM96,BMS97}.
Individuals have disease-causing contacts only along the links in this
network.  We distinguish here between ``connections'' and actual contacts.
Connections between pairs of individuals predispose those individuals to
disease-causing contact, but do not guarantee it.  An individual's
connections are the set of people with whom the individual may have contact
during the time he or she is infective---people that the individual lives
with, works with, sits next to on the bus, and so forth.

We can vary the number of connections each person has with others by
choosing a particular degree distribution for the network.  (Recall that
the degree of a vertex in a network is the number of other vertices to
which it is attached.)  For example, in the case of sexual contacts, which
can transmit STDs, the degree distribution has been found to follow a
power-law form~\cite{Liljeros01}.  By placing the model on a network with a
power-law degree distribution we can emulate this effect in our model.

Our second modification of the model is to allow the probability of
disease-causing contact between pairs of individuals who have a connection
to vary, so that some pairs have higher probability of disease transmission
than others.

Consider a pair of individuals who are connected, one of whom $i$ is
infective and the other $j$ susceptible.  Suppose that the average rate of
disease-causing contacts between them is $r_{ij}$, and that the infective
individual remains infective for a time~$\tau_i$.  Then the probability
$1-T_{ij}$ that the disease will {\em not\/} be transmitted from $i$ to $j$
is
\begin{equation}
1 - T_{ij} = \lim_{\delta t\to0} (1-r_{ij}\delta t)^{\tau_i/\delta t}
           = \e^{-r_{ij}\tau_i},
\label{deltat}
\end{equation}
and the probability of transmission is
\begin{equation}
T_{ij} = 1 - \e^{-r_{ij}\tau_i}.
\label{defstij}
\end{equation}
Some models, particularly computer simulations, use discrete time-steps
rather than continuous time, in which case instead of taking the limit in
Eq.~\eref{deltat} we simply set $\delta t=1$, giving
\begin{equation}
T_{ij} = 1 - (1-r_{ij})^{\tau_i},
\label{defstijdisc}
\end{equation}
where $\tau$ is measured in time-steps.

In general $r_{ij}$ and $\tau_i$ will vary between individuals, so that the
probability of transmission also varies.  Let us assume initially that
these two quantities are iid random variables chosen from some appropriate
distributions $P(r)$ and $P(\tau)$.  (We will relax this assumption later.)
The rate $r_{ij}$ need not be symmetric---the probability of transmission
in either direction might not be the same.  In any case, $T_{ij}$ is in
general not symmetric because of the appearance of $\tau_i$ in
Eqs.~\eref{defstij} and~\eref{defstijdisc}.

Now here's the trick: because $r_{ij}$ and $\tau_i$ are iid random
variables, so is $T_{ij}$, and hence the {\it a priori\/} probability of
transmission of the disease between two individuals is simply the average
$T$ of $T_{ij}$ over the distributions $P(r)$ and $P(\tau)$, which is
\begin{equation}
T = \av{T_{ij}} = 1 - \int_0^\infty \d r\>\d\tau \> P(r) P(\tau)\,\e^{-r\tau}
\label{defstcont}
\end{equation}
for the continuous time case or
\begin{equation}
T = 1 - \int_0^\infty \d r\>\sum_{\tau=0}^\infty \> P(r) P(\tau)\,(1-r)^\tau
\label{defstdisc}
\end{equation}
for the discrete case~\cite{WSS01}.  We call $T$ the ``transmissibility''
of the disease.  It is necessarily always in the range $0\le T\le1$.

Thus the fact that individual transmission probabilities vary makes no
difference whatsoever; in the population as a whole the disease will
propagate as if all transmission probabilities were equal to~$T$.  We
demonstrate the truth of this result by explicit simulation in
Section~\ref{example}.  It is this result that makes our models solvable.
Cases in which the variables $r$ and $\tau$ are not iid are trickier, but,
as we will show, these are sometimes solvable as well.

We note further that more complex disease transmission models, such as SEIR
models in which there is an infected-but-not-infective period (E), are also
covered by this formalism.  The transmissibility $T_{ij}$ is essentially
just the integrated probability of transmission of the disease between two
individuals.  The precise temporal behavior of infectivity and other
variables is unimportant.  Indeed the model can be generalized to include
{\em any\/} temporal variation in infectivity of the infective individuals,
and transmission can still be represented correctly by a simple
transmissibility variable~$T$, as above.

Now imagine watching an outbreak of the disease, which starts with a single
infective individual, spreading across our network.  If we were to mark or
``occupy'' each edge in the graph across which the disease is transmitted,
which happens with probability $T$, the ultimate size of the outbreak would
be precisely the size of the cluster of vertices that can be reached from
the initial vertex by traversing only occupied edges.  Thus, the model is
precisely equivalent to a bond percolation model with bond occupation
probability $T$ on the graph representing the community.  The connection
between the spread of disease and percolation was in fact one of the
original motivations for the percolation model itself~\cite{FH63}, but
seems to have been formulated in the manner presented here first by
Grassberger~\cite{Grassberger83} for the case of uniform $r$ and $\tau$,
and by Warren~\etal~\cite{WSS01,Warren02} for the non-uniform case.

In the next section we show how the percolation problem can be solved on
random graphs with arbitrary degree distributions, giving exact solutions
for the typical size of outbreaks, presence of an epidemic, size of the
epidemic (if there is one), and a number of other quantities of interest.

\section{Exact solutions on networks with arbitrary degree distributions}
\label{simplest}
One of the most important results to come out of empirical work on networks
is the finding that the degree distributions of many networks are highly
right-skewed.  In other words, most vertices have only a low degree, but
there are a small number whose degree is very
high~\cite{Price65,AJB99,ASBS00,AB02}.  The network of sexual contacts
discussed above provides one example of such a
distribution~\cite{Liljeros01}.  It is known that the presence of highly
connected vertices can have a disproportionate effect on certain properties
of the network.  Recent work suggests that the same may be true for disease
propagation on networks~\cite{PV01a,LM01}, and so it will be important that
we incorporate non-trivial degree distributions in our models.  As a first
illustration of our method therefore, we look at a simple class of
unipartite graphs studied previously by a variety of
authors~\cite{BC78,MR95,MR98,NSW01,NWS02,CEBH00,CEBH01,CNSW00,DMS01a,ALPH01},
in which the degree distribution is specified, but the graph is in other
respects random.

Our graphs are simply defined.  One specifies the degree distribution by
giving the properly normalized probabilities $p_k$ that a randomly chosen
vertex has degree~$k$.  A set of $N$ degrees $\set{k_i}$, also called a
degree sequence, is then drawn from this distribution and each of the $N$
vertices in the graph is given the appropriate number $k_i$ of
``stubs''---ends of edges emerging from it.  Pairs of these stubs are then
chosen at random and connected together to form complete edges.  Pairing of
stubs continues until none are left.  (If an odd number of stubs is by
chance generated, complete pairing is not possible, in which case we
discard one $k_i$ and draw another until an even number is achieved.)  This
technique guarantees that the graph generated is chosen uniformly at random
from the set of all graphs with the selected degree sequence.

All the results given in this section are averaged over the ensemble of
possible graphs generated in this way, in the limit of large graph size.

\subsection{Generating functions}
We wish then to solve for the average behavior of graphs of this type under
bond percolation with bond occupation probability $T$.  We will do this
using generating function techniques~\cite{Wilf94}.  Following
Newman~\etal~\cite{NSW01}, we define a generating function for the degree
distribution thus:
\begin{equation}
G_0(x) = \sum_{k=0}^\infty p_k x^k.
\label{defsg0}
\end{equation}
Note that $G_0(1)=\sum_k p_k=1$ if $p_k$ is a properly normalized
probability distribution.

This function encapsulates all of the information about the degree
distribution.  Given it, we can easily reconstruct the distribution by
repeated differentiation:
\begin{equation}
p_k = {1\over k!} {\d^k G_0\over\d x^k}\bigg|_{x=0}.
\label{derivatives}
\end{equation}
We say that the generating function $G_0$ ``generates'' the
distribution~$p_k$.  The generating function is easier to work with than
the degree distribution itself because of two crucial properties:

\medbreak\noindent {\bf Powers}\quad If the distribution of a property $k$
of an object is generated by a given generating function, then the
distribution of the sum of $k$ over $m$ independent realizations of the
object is generated by the $m^{\rm th}$ power of that generating function.
For example, if we choose $m$ vertices at random from a large graph, then
the distribution of the sum of the degrees of those vertices is generated
by~$[G_0(x)]^m$.

\medbreak\noindent {\bf Moments}\quad The mean of the probability
distribution generated by a generating function is given by the first
derivative of the generating function, evaluated at~1.  For instance, the
mean degree $z$ of a vertex in our network is given by
\begin{equation}
z = \av{k} = \sum_k k p_k = G_0'(1).
\label{avk}
\end{equation}
Higher moments of the distribution can be calculated from higher
derivatives also.  In general, we have
\begin{equation}
\av{k^n} = \sum_k k^n p_k = 
  \biggl[ \biggl(x {\d\over\d x}\biggr)^{\!n} G_0(x) \biggr]_{x=1}.
\end{equation}

A further observation that will also prove crucial is the following.  While
$G_0$ above correctly generates the distribution of degrees of randomly
chosen vertices in our graph, a different generating function is needed for
the distribution of the degrees of vertices reached by following a randomly
chosen edge.  If we follow an edge to the vertex at one of its ends, then
that vertex is more likely to be of high degree than is a randomly chosen
vertex, since high-degree vertices have more edges attached to them than
low-degree ones.  The distribution of degrees of the vertices reached by
following edges is proportional to $kp_k$, and hence the generating
function for those degrees is
\begin{equation}
{\sum_k k p_k x^k\over\sum_k k p_k} = x {G_0'(x)\over G_0'(1)}.
\label{arrival}
\end{equation}
In general we will be concerned with the number of ways of leaving such a
vertex \emph{excluding} the edge we arrived along, which is the degree
minus~1.  To allow for this, we simply divide the function above by one
power of $x$, thus arriving at a new generating function
\begin{equation}
G_1(x) = {G_0'(x)\over G_0'(1)} = {1\over z}\, G_0'(x),
\label{defsg1}
\end{equation}
where $z$ is the average vertex degree, as before.

In order to solve the percolation problem, we will also need generating
functions $G_0(x;T)$ and $G_1(x;T)$ for the distribution of the number of
\emph{occupied} edges attached to a vertex, as a function of the
transmissibility~$T$.  These are simple to derive.  The probability of a
vertex having exactly $m$ of the $k$ edges emerging from it occupied is
given by the binomial distribution ${k\choose m} T^m (1-T)^{k-m}$, and
hence the probability distribution of $m$ is generated by
\begin{eqnarray}
G_0(x;T) &=& \sum_{m=0}^\infty \sum_{k=m}^\infty p_k
           \biggl({k\atop m}\biggr) T^m (1-T)^{k-m} x^m\nonumber\\
         &=& \sum_{k=0}^\infty p_k \sum_{m=0}^k
           \biggl({k\atop m}\biggr) (xT)^m (1-T)^{k-m}\nonumber\\
         &=& \sum_{k=0}^\infty p_k (1-T+xT)^k\nonumber\\
         &=& G_0(1+(x-1)T).
\label{defsg0xt}
\end{eqnarray}
Similarly, the probability distribution of occupied edges leaving a vertex
arrived at by following a randomly chosen edge is generated by
\begin{equation}
G_1(x;T) = G_1(1+(x-1)T).
\label{defsg1xt}
\end{equation}
Note that, in our notation
\begin{subequations}
\begin{eqnarray}
G_0(x;1)  &=& G_0(x),\\
G_0(1;T)  &=& G_0(1),\\
G_0'(1;T) &=& TG_0'(1),
\end{eqnarray}
\end{subequations}
and similarly for $G_1$.  ($G_0'(x;T)$ here represents the derivative of
$G_0(x;T)$ with respect to its first argument.)

\subsection{Outbreak size distribution}
\label{percolation}
The first quantity we will work out is the distribution $P_s(T)$ of the
sizes $s$ of outbreaks of the disease on our network, which is also the
distribution of sizes of clusters of vertices connected together by
occupied edges in the corresponding percolation model.  Let $H_0(x;T)$ be
the generating function for this distribution:
\begin{equation}
H_0(x;T) = \sum_{s=0}^\infty P_s(T) x^s.
\end{equation}
By analogy with the previous section we also define $H_1(x;T)$ to be the
generating function for the cluster of connected vertices we reach by
following a randomly chosen edge.

Now, following Ref.~\onlinecite{NSW01}, we observe that $H_1$ can be broken
down into an additive set of contributions as follows.  The cluster reached
by following an edge may be:
\begin{enumerate}
\item a single vertex with no occupied edges attached to it, other than the
  one along which we passed in order to reach it;
\item a single vertex attached to any number $m\ge1$ of occupied edges
other than the one we reached it by, each leading to another cluster whose
size distribution is also generated by~$H_1$.
\end{enumerate}
We further note that the chance that any two finite clusters that are
attached to the same vertex will have an edge connecting them together
directly goes as $N^{-1}$ with the size~$N$ of the graph, and hence is zero
in the limit $N\to\infty$.  In other words, there are no loops in our
clusters; their structure is entirely tree-like.

Using these results, we can express $H_1(x;T)$ in a Dyson-equation-like
self-consistent form thus:
\begin{equation}
H_1(x;T) = x G_1(H_1(x;T);T).
\label{bondh1}
\end{equation}
Then the size of the cluster reachable from a randomly chosen starting
vertex is distributed according to
\begin{equation}
H_0(x;T) = x G_0(H_1(x;T);T).
\label{bondh0}
\end{equation}
It is straightforward to verify that for the special case $T=1$ of 100\%
transmissibility, these equations reduce to those given in
Ref.~\onlinecite{NSW01} for component size in random graphs with arbitrary
degree distributions.  Equations~\eref{bondh1} and~\eref{bondh0} provide
the solution for the more general case of finite transmissibility which
applies to SIR models.  Once we have $H_0(x;T)$, we can extract the
probability distribution of clusters $P_s(T)$ by differentiation using
Eq.~\eref{derivatives} on~$H_0$.  In most cases however it is not possible
to find arbitrary derivatives of $H_0$ in closed form.  Instead we
typically evaluate them numerically.  Since direct evaluation of numerical
derivatives is prone to machine precision problems, we recommend evaluating
the derivatives by numerical contour integration using the Cauchy formula:
\begin{equation}
P_s(T) = {1\over s!}\,{\d^s\!H_0\over\d x^s}\bigg|_{x=0}
       = {1\over2\pi\i} \oint {H_0(\zeta;T)\over \zeta^{s+1}} \d\zeta,
\label{cauchy}
\end{equation}
where the integral is over the unit circle~\cite{MN00b}.  It is possible to
find the first thousand derivatives of a function without difficulty using
this method~\cite{NSW01}.  By this method then, we can find the exact
probability $P_s$ that a particular outbreak of our disease will infect $s$
people in total, as a function of the transmissibility~$T$.

\subsection{Outbreak sizes and the epidemic transition}
Although in general we must use numerical methods to find the complete
distribution $P_s$ of outbreak sizes from Eq.~\eref{cauchy}, we can
find the mean outbreak size in closed form.  Using Eq.~\eref{avk}, we
have
\begin{equation}
\av{s} = H_0'(1;T) = 1 + G_0'(1;T) H_1'(1;T),
\end{equation}
where we have made use of the fact that the generating functions are 1 at
$x=1$ if the distributions that they generate are properly normalized.
Differentiating Eq.~\eref{bondh1}, we have
\begin{equation}
H_1'(1;T) = 1 + G_1'(1;T) H_1'(1;T) = {1\over1-G_1'(1;T)},
\end{equation}
and hence
\begin{equation}
\av{s} = 1 + {G_0'(1;T)\over1 - G_1'(1;T)}
       = 1 + {T G_0'(1)\over1 - T G_1'(1)}.
\label{avs}
\end{equation}
Given Eqs.~\eref{defsg0}, \eref{defsg1}, \eref{defsg0xt},
and~\eref{defsg1xt}, we can then evaluate this expression to get the mean
outbreak size for any value of $T$ and degree distribution.

We note that Eq.~\eref{avs} diverges when $T G_1'(1)=1$.  This point marks
the onset of an epidemic; it is the point at which the typical outbreak
ceases to be confined to a finite number of individuals, and expands to
fill an extensive fraction of the graph.  The transition takes place when
$T$ is equal to the critical transmissibility $T_c$, given by
\begin{equation}
T_c = {1\over G_1'(1)} = {G_0'(1)\over G_0''(1)}
    = {\sum_k k p_k\over\sum_k k(k-1) p_k}.
\label{tc}
\end{equation}

For $T>T_c$, we have an epidemic, or ``giant component'' in the language of
percolation.  We can calculate the size of this epidemic as follows.  Above
the epidemic threshold Eq.~\eref{bondh1} is no longer valid because the
giant component is extensive and therefore can contain loops, which
destroys the assumptions on which Eq.~\eref{bondh1} was based.  The
equation \emph{is} valid however if we redefine $H_0$ to be the generating
function only for outbreaks other than epidemic outbreaks, i.e.,~isolated
clusters of vertices that are not connected to the giant component.  These
however do not fill the entire graph, but only the portion of it not
affected by the epidemic.  Thus, above the epidemic transition, we have
\begin{equation}
H_0(1;T) = \sum_s P_s = 1 - S(T),
\label{giant1}
\end{equation}
where $S(T)$ is the fraction of the population affected by the epidemic.
Rearranging Eq.~\eref{giant1} for~$S$ and making use of Eq.~\eref{bondh0},
we find that the size of the epidemic is
\begin{equation}
S(T) = 1 - G_0(u;T),
\label{giant2}
\end{equation}
where $u\equiv H_1(1;T)$ is the solution of the self-consistency relation
\begin{equation}
u = G_1(u;T).
\label{giant3}
\end{equation}
Results equivalent to Eqs.~\eref{avs} to~\eref{giant3} were given
previously in a different context in Ref.~\onlinecite{CNSW00}.

Note that it is not the case, even above $T_c$, that all outbreaks give
rise to epidemics of the disease.  There are still finite outbreaks even in
the epidemic regime.  While this appears very natural, it stands
nonetheless in contrast to the standard fully mixed models, for which all
outbreaks give rise to epidemics above the epidemic transition point.  In
the present case, the probability of an outbreak becoming an epidemic at a
given $T$ is simply equal to~$S(T)$.

\subsection{Degree of infected individuals}
The quantity~$u$ defined in Eq.~\eref{giant3} has a simple interpretation:
it is the probability that the vertex at the end of a randomly chosen edge
remains uninfected during an epidemic (i.e.,~that it belongs to one of the
finite components).  The probability that a vertex becomes infected via one
of its edges is thus $v=1-T+Tu$, which is the sum of the probability $1-T$
that the edge is unoccupied, and the probability $Tu$ that it is occupied
but connects to an uninfected vertex.  The total probability of being
uninfected if a vertex has degree~$k$ is~$v^k$, and the probability of
having degree~$k$ given that a vertex is uninfected is $p_kv^k/\sum_k
p_kv^k=p_k v^k/G_0(v)$, which distribution is generated by the function
$G_0(vx)/G_0(v)$.  Differentiating and setting $x=1$, we then find that the
average degree $z_{\rm out}$ of vertices outside the giant component is
\begin{equation}
z_{\rm out} = {v G_0'(v)\over G_0(v)} = {v G_1(v)\over G_0(v)}\,z
            = {u[1-T+Tu]\over1-S}\,z.
\label{zout}
\end{equation}
Similarly the degree distribution for an infected vertex is generated by
$[G_0(x)-G_0(vx)]\big/[1-G_0(v)]$, which gives a mean degree~$z_{\rm in}$
for vertices in the giant component of
\begin{equation}
z_{\rm in} = {1-vG_1(v)\over 1-G_0(v)}\,z = {1 - u[1-T+Tu]\over S}\,z.
\end{equation}

Note that $1-S=G_0(u;T)\le u$, since all coefficients of $G_0(x;T)$ are by
definition positive (because they form a probability distribution) and
hence $G_0(x;T)$ has only positive derivatives, meaning that it is convex
everywhere on the positive real line within its domain of convergence.
Thus, from Eq.~\eref{zout}, $z_{\rm out}\le z$.  Similarly, $z_{\rm in}\ge
z$, and hence, as we would expect, the mean degree of infected individuals
is always greater than or equal to the mean degree of uninfected ones.
Indeed, the probability of a vertex being infected, given that it has
degree~$k$, goes as $1-v^k=1-\e^{-k\log(1/v)}$, i.e.,~tends exponentially
to unity as degree becomes large.

\subsection{An example}
\label{example}
Let us now look at an application of these results to a specific example of
disease spreading.  First of all we need to define our network of
connections between individuals, which means choosing a degree
distribution.  Here we will consider graphs with the degree distribution
\begin{equation}
p_k = \biggl\lbrace \begin{array}{ll}
            0                                      & \mbox{for $k=0$}\\
            C k^{-\alpha} \e^{-k/\kappa} \qquad\null & \mbox{for $k\ge1$.}
      \end{array}
\end{equation}
where $C$, $\alpha$, and $\kappa$ are constants.  In other words, the
distribution is a power-law of exponent $\alpha$ with an exponential cutoff
around degree~$\kappa$.  This distribution has been studied before by
various authors~\cite{ASBS00,NSW01,NWS02,CNSW00}.  It makes a good example
for a number of reasons:
\begin{enumerate}
\item distributions of this form are seen in a variety of real-world
  networks~\cite{ASBS00,Newman01a};
\item it includes pure power-law and pure exponential distributions, both
  of which are also seen in various
  networks~\cite{Price65,AJB99,FFF99,ASBS00}, as special cases when
  $\kappa\to\infty$ or $\alpha\to0$;
\item it is normalizable and has all moments finite for any finite~$\kappa$;
\end{enumerate}

The constant $C$ is fixed by the requirement of normalization, which gives
$C = [\Li_\alpha(\e^{-1/\kappa})]^{-1}$ and hence
\begin{equation}
p_k = {k^{-\alpha} \e^{-k/\kappa}\over\Li_\alpha(\e^{-1/\kappa})} \qquad
\mbox{for $k\ge1$,}
\label{powerlaw}
\end{equation}
where $\Li_n(x)$ is the $n$th polylogarithm of $x$.

We also need to choose the distributions $P(r)$ and $P(\tau)$ for the
transmission rate and the time spent in the infective state.  For the sake
of easier comparison with computer simulations we use discrete time and
choose both distributions to be uniform, with $r$ real in the range $0\le
r<r_{\rm max}$ and $\tau$ integer in the range $1\le\tau\le\tau_{\rm max}$.
The transmissibility~$T$ is then given by Eq.~\eref{defstdisc}.  From
Eq.~\eref{powerlaw}, we have
\begin{equation}
G_0(x) = {\Li_\alpha(x\e^{-1/\kappa})\over\Li_\alpha(\e^{-1/\kappa})}.
\end{equation}
and
\begin{equation}
G_1(x) = {\Li_{\alpha-1}(x\e^{-1/\kappa})\over
          x\Li_{\alpha-1}(\e^{-1/\kappa})}.
\end{equation}
Thus the epidemic transition in this model occurs at
\begin{equation}
T_c = {\Li_{\alpha-1}(\e^{-1/\kappa})\over
       \Li_{\alpha-2}(\e^{-1/\kappa})-\Li_{\alpha-1}(\e^{-1/\kappa})}.
\end{equation}
Below this value of $T$ there are only small (non-epidemic) outbreaks,
which have mean size
\begin{eqnarray}
\av{s} &=& 1 + \nonumber\\
       & & \hspace{-10mm} {T[\Li_{\alpha-1}(\e^{-1/\kappa})]^2\over
          \Li_\alpha(\e^{-1/\kappa})
          [(T+1)\Li_{\alpha-1}(\e^{-1/\kappa})
           - T\Li_{\alpha-2}(\e^{-1/\kappa})]}.\nonumber\\
\end{eqnarray}
Above it, we are in the region in which epidemics can occur, and they
affect a fraction $S$ of the population in the limit of large graph size.
We cannot solve for $S$ in closed form, but we can solve
Eqs.~\eref{giant2} and~\eref{giant3} by numerical iteration and hence
find~$S$.

\begin{figure}
\begin{center}
\resizebox{\figurewidth}{!}{\includegraphics{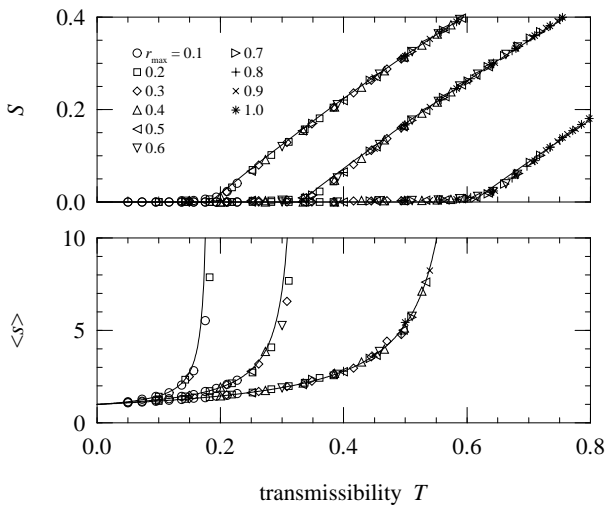}}
\end{center}
\caption{Epidemic size (top) and average outbreak size (bottom) for the SIR
  model on networks with degree distributions of the form~\eref{powerlaw}
  as a function of transmissibility.  Solid lines are the exact solutions,
  Eqs.~\eref{giant2} and~\eref{avs}, for $\alpha=2$ and (left to right in
  each panel) $\kappa=20$, 10, and~5.  Each of the points is an average
  result for $10\,000$ simulations on graphs of $100\,000$ vertices each
  with distributions of $r$ and~$\tau$ as described in the text.}
\label{siruni}
\end{figure}

In Fig.~\ref{siruni} we show the results of calculations of the average
outbreak size and the size of epidemics from the exact formulas, compared
with explicit simulations of the SIR model on networks with the degree
distribution~\eref{powerlaw}.  Simulations were performed on graphs of
$N=100\,000$ vertices, with $\alpha=2$, a typical value for networks seen
in the real world, and $\kappa=5$, 10, and 20 (the three curves in each
panel of the figure).  For each pair of the parameters $\alpha$ and
$\kappa$ for the network, we simulated $10\,000$ disease outbreaks each for
$(r,\tau)$ pairs with $r_{\rm max}$ from 0.1 to 1.0 in steps of $0.1$, and
$\tau_{\rm max}$ from 1 to~10 in steps of~1.  Fig.~\ref{siruni} shows all
of these results on one plot as a function of the transmissibility~$T$,
calculated from Eq.~\eref{defstdisc}.

The figure shows two important things.  First, the points corresponding to
different values of $r_{\rm max}$ and $\tau_{\rm max}$ but the same value
of $T$ fall in the same place and the two-parameter set of results for $r$
and $\tau$ collapses onto a single curve.  This indicates that the
arguments leading to Eqs.~\eref{defstcont} and~\eref{defstdisc} are correct
(as also demonstrated by Warren~\etal~\cite{WSS01,Warren02}) and that the
statistical properties of the disease outbreaks really do depend only on
the transmissibility~$T$, and not on the individual rates and times of
infection.  Second, the data clearly agree well with our analytic results
for average outbreak size and epidemic size, confirming the correctness of
our exact solution.  The small disagreement between simulations and exact
solution for $\av{s}$ close to the epidemic transition in the lower panel
of the figure appears to be a finite size effect, due to the relatively
small system sizes used in the simulations.

To emphasize the difference between our results and those for the
equivalent fully mixed model, we compare the position of the epidemic
threshold in the two cases.  In the case $\alpha=2$, $\kappa=10$ (the
middle curve in each frame of Fig.~\ref{siruni}), our analytic solution
predicts that the epidemic threshold occurs at $T_c=0.329$.  The
simulations agree well with this prediction, giving $T_c=0.32(2)$.  By
contrast, a fully mixed SIR model in which each infective individual
transmits the disease to the same average number of others as in our
network, gives a very different prediction of $T_c=0.558$.

\section{Correlated transmission probabilities}
\label{correlated}
It is possible to imagine many cases in which the probabilities of
transmission of a disease from an infective individual to those with whom
he or she has connections are not iid random variables.  In other words,
the probabilities of transmission from a given individual to others could
be drawn from different distributions for different individuals.  This
allows, for example, for cases in which the probabilities tend either all
to be high or all to be low but are rarely a mixture of the two.  In this
section, we show how the model of Section~\ref{simplest} can be generalized
to allow for this.

Suppose that the transmission rates $r$ for transmission from an infective
individual~$i$ to each of the $k_i$ others with whom they have connections
are drawn from a distribution~$P_i(r)$, which can vary from one individual
to another in any way we like.  Thus the \textit{a priori} probability of
transmission from~$i$ to any one of his or her neighbors in the network is
\begin{equation}
T_i = 1 - \int_0^\infty \d r\>\d\tau \> P_i(r) P(\tau)\, \e^{-r\tau}.
\label{defsti}
\end{equation}
One could of course also allow the distribution from which the time $\tau$
is drawn to vary from one individual to another, although this doesn't
result in any functional change in the theory, so it would be rather
pointless.  In any case, the formalism developed here can handle this type
of dependency perfectly well.

Following Eq.~\eref{defsg0xt}, we note that in the percolation
representation of our model the distribution of the number of occupied
edges leading from a particular vertex is now generated by the function
\begin{eqnarray}
G_0(x;\set{T_i}) &=& {1\over N}\sum_{i=0}^N \sum_{m=0}^{k_i}
              \biggl({k_i\atop m}\biggr) T_i^m (1-T_i)^{k_i-m} x^m\nonumber\\
                 &=& {1\over N}\sum_{i=0}^N (1+(x-1)T_i)^{k_i}.
\label{g0ti}
\end{eqnarray}
And similarly, the probability distribution of occupied edges leaving a
vertex arrived at by following a randomly chosen edge is generated by
\begin{equation}
G_1(x;\set{T_i}) = {\sum_i k_i (1+(x-1)T_i)^{k_i-1}\over\sum_i k_i}.
\label{g1ti}
\end{equation}
Clearly these reduce to Eqs.~\eref{defsg0xt} and~\eref{defsg1xt} when $T_i$
is independent of~$i$.

With these definitions of the basic generating functions, our derivations
proceed as before.  The complete distribution of the sizes of outbreaks of
the disease, excluding epidemic outbreaks if there are any, is generated by
\begin{equation}
H_0(x;\set{T_i}) = x G_0(H_1(x;\set{T_i});\set{T_i}),
\label{corrh0}
\end{equation}
where
\begin{equation}
H_1(x;\set{T_i}) = x G_1(H_1(x;\set{T_i});\set{T_i}).
\label{corrh1}
\end{equation}
The average outbreak size when there is no epidemic is given by
Eq.~\eref{avs} as before, and the size of epidemics above the epidemic
transition is given by Eqs.~\eref{giant2} and~\eref{giant3}.  The
transition itself occurs when $G_1'(1;\set{T_i})=1$ and, substituting for
$G_1$ from Eq.~\eref{g1ti}, we can also write this in the form
\begin{equation}
\sum_{i=0}^N k_i[(k_i-1)T_i-1] = 0.
\label{ticondition}
\end{equation}
In fact, it is straightforward to convince oneself that when the sum on the
left-hand side of this equation is greater than zero epidemics occur, and
when it is less than zero they do not.

For example, consider the special case in which the distribution of
transmission rates $P(r)$ depends on the degree of the vertex representing
the infective individual.  One could imagine, for example, that individuals
with a large number of connections to others tend to have lower
transmission rates than those with only a small number.  In this case $T_i$
is a function only of $k_i$ and hence we have
\begin{eqnarray}
G_0(x;\set{T_k}) &=& {1\over N} \sum_{i=0}^N
                     (1+(x-1)T_{k_i})^{k_i}\nonumber\\
                 &=& \sum_{k=0}^\infty p_k (1+(x-1)T_k)^k,
\label{g0tk}
\end{eqnarray}
and
\begin{equation}
G_1(x;\set{T_k}) = {\sum_k k p_k (1+(x-1)T_k)^{k-1}\over\sum_k k p_k},
\label{g1tk}
\end{equation}
where $T_k$ is the mean transmissibility for vertices of degree~$k$.

One can also treat the case in which the transmissibility is a function of
the number of connections which the individual being infected has.  If the
probability of transmission to an individual with degree $k$ is $U_k$, then
we define
\begin{eqnarray}
\label{g0uk}
G_0(x;\set{U_k}) &=& \sum_k p_k x^k,\\
\label{g1uk}
G_1(x;\set{U_k}) &=& {\sum_k k p_k [1 + (x^{k-1}-1) U_k]\over\sum_k k p_k},
\end{eqnarray}
and then the calculation of cluster size distribution and so forth proceeds
as before.

Further, one can solve the case in which probability of transmission of the
disease depends on {\em both\/} the probabilities of giving it and catching
it, which are arbitrary functions $T_k$ and $U_k$ of the numbers of
connections of the infective and susceptible individuals.  (This means that
transmission from a vertex with degree~$j$ to a vertex with degree~$k$
occurs with a probability equal to the product $T_jU_k$.)  The appropriate
generating functions for this case are
\begin{eqnarray}
\label{g0tkuk}
G_0(x;\set{T_k},\set{U_k}) &=& \sum_k p_k (1+(x-1)T_k)^k,\\
\label{g1tkuk}
G_1(x;\set{T_k},\set{U_k}) &=& \nonumber\\
& & \hspace{-28mm}
    {\sum_k kp_k [1+((1+(x-1)T_k)^{k-1}-1)U_k]\over\sum_k k p_k},
\end{eqnarray}
and indeed Eqs.~\eref{g0tk} to~\eref{g1uk} can be viewed as special cases
of these equations when either $T_k=1$ or $U_k=1$ for all~$k$.  Note that
$G_0(x;\set{U_k})$ and $G_0(x;\set{T_k},\set{U_k})$ are both independent of
$\set{U_k}$, since the probability of a randomly-chosen infective
individual having the disease is unity, regardless of the probability that
they caught it in the first place.

As a concrete example of the developments of this section, consider the
physically plausible case in which the transmissibility~$T$ depends
inversely on the degree of the infective individual: $T_k=T_1/k$.  Then
from Eq.~\eref{ticondition} we find that there is epidemic behavior only if
\begin{equation}
T_1 > {z\over z-1},
\end{equation}
regardless of the degree distribution.  Since $T$ lies strictly between
zero and one however, this is impossible.  In networks of this type, we
therefore conclude that diseases cannot spread.  Only if transmissibilities
fall off slower than inversely with degree in at least some part of their
range are epidemics possible.  One plausible way in which this might happen
is if $T_k\sim(T_0+k)^{-1}$.  In this case it is straightforward to show
that epidemics are possible for some degree distributions for some values
of~$T_0$.

Other extensions of the model are possible too.  One area of current
interest is models incorporating vaccination~\cite{MN00a,PV02}.  Disease
propagation on networks incorporating vaccinated individuals can be
represented as a joint site/bond percolation process, which can also be
solved exactly~\cite{CNSW00}, both in the case of uniform independent
vaccination probability (i.e.,~random vaccination of a population) and in
the case of vaccination that is correlated with properties of individuals
such as their degree (so that vaccination can be directed at the so-called
core group of the disease-carrying network---those with the highest
degrees).

\section{Structured populations}
\label{structured}
The models we have studied so far have made use of simple unipartite graphs
as the substrate for the spread of disease.  These graphs may have any
distribution we choose of the degrees of their vertices, but in all other
respects are completely random.  Many of the really interesting cases of
disease spreading take place on networks that have more structure than
this.  Cases that have been studied previously include disease spreading
among children who attend a common school and among patients in different
wards of a hospital between whom pathogens are communicated by peripatetic
caregivers~\cite{Hyde01}.  Here, we give just one example of disease
spreading in a population with a very simple structure.  The example we
consider is the spread of a sexually transmitted disease.  The important
structural element of the population in this case is its division into men
and women.

\begin{figure}
\resizebox{\figurewidth}{!}{\includegraphics{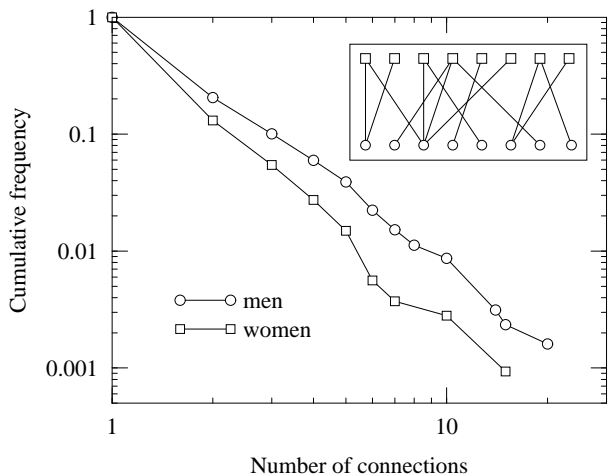}}
\caption{Distributions of the numbers of sexual contacts of men and women
  in the study of Liljeros~\etal~\protect\cite{Liljeros01}.  The histogram
  is cumulative, meaning that the vertical axis indicates the fraction of
  individuals studied who have greater than or equal to the number of
  contacts specified on the horizontal axis.  Both distributions
  approximately follow power laws---straight lines on the logarithmic axes
  used here.  Inset: the bipartite form of the modeled network of
  contacts.}
\label{menwomen}
\end{figure}

\subsection{Bipartite populations}
Consider then a population of $M$ men and $N$ women, who have distributions
$p_j$, $q_k$ of their numbers $j$ and $k$ of possibly disease-causing
contacts with the opposite sex (connections in our nomenclature).  In a
recent study of 2810 respondents Liljeros~\etal~\cite{Liljeros01} recorded
the numbers of sexual partners of men and women over the course of a year
and found the distributions $p_j$, $q_k$ shown in Fig.~\ref{menwomen}.  As
the figure shows, the distributions appear to take a power-law form
$p_j\sim j^{\alpha_m}$, $q_k\sim k^{\alpha_f}$, with exponents $\alpha_m$
and $\alpha_f$ that fall in the range $3.1$ to $3.3$ for both men and
women~\footnote{One should observe that the network studied in
  Ref.~\onlinecite{Liljeros01} is a cumulative network of actual sexual
  contacts---it represents the sum of all contacts over a specified period
  of time.  Although this is similar to other networks of sexual contacts
  studied previously~\cite{GAM89,Klovdahl94}, it is not the network
  required by our models, which is the instantaneous network of connections
  (not contacts---see Section~\ref{perc}).  While the network measured may
  be a reasonable proxy for the network we need, it is not known if this is
  the case.}.  (The exponent for women seems to be a little higher than
that for men, but the difference is smaller than the statistical error on
the measurement.)

We will assume that the disease of interest is transmitted primarily by
contacts between men and women (true only for some diseases in some
communities~\cite{HY84}), so that, to a good approximation, the network of
contacts is bipartite, as shown in the inset of Fig.~\ref{menwomen}.  That
is, there are two types of vertices representing men and women, and edges
representing connections run only between vertices of unlike kinds.  With
each edge we associate two transmission rates, one of which represents the
probability of disease transmission from male to female, and the other from
female to male.  These rates are drawn from appropriate distributions as
before, as are the times for which men and women remain infective.  Also as
before, however, it is only the average integrated probability of
transmission in each direction that matters for our percolation model, so
that we have two transmissibilities $T_{mf}$ and $T_{fm}$ for the two
directions~\footnote{It is also worth noting that networks of sexual
  contacts observed in sociometric studies~\cite{Klovdahl94} are often
  highly dendritic, with few short loops, indicating that the tree-like
  components of our percolating clusters may be, at least in this respect,
  quite a good approximation to the shape of real STD outbreaks.}.

We define two pairs of generating functions for the degree distributions of
males and females:
\begin{subequations}
\begin{eqnarray}
f_0(x) &=& \sum_j p_j x^j,\qquad         f_1(x) = \frac{1}{\mu} f_0'(x),\\
g_0(x) &=& \sum_k q_k x^k,\hspace{6.8mm} g_1(x) = \frac{1}{\nu} g_0'(x),
\end{eqnarray}
\end{subequations}
where $\mu$ and $\nu$ are the averages of the two degree distributions, and
are related by
\begin{equation}
{\mu\over M} = {\nu\over N},
\end{equation}
since the total numbers of edges ending at male and female vertices are
necessarily the same.  Using these functions we further define, as before
\begin{subequations}
\begin{eqnarray}
f_0(x;T) &=& f_0(1+(x-1)T),\\
f_1(x;T) &=& f_1(1+(x-1)T),\\
g_0(x;T) &=& g_0(1+(x-1)T),\\
g_1(x;T) &=& g_1(1+(x-1)T).
\end{eqnarray}
\end{subequations}

Now consider an outbreak that starts at a single individual, who for the
moment we take to be male.  From that male the disease will spread to some
number of females, and from them to some other number of males, so that
after those two steps a number of new males will have contracted the
disease, whose distribution is generated by
\begin{equation}
F_0(x;T_{mf},T_{fm}) = f_0(g_1(x;T_{fm});T_{mf}).
\end{equation}
For a disease arriving at a male vertex along a randomly chosen edge we
similarly have
\begin{equation}
F_1(x;T_{mf},T_{fm}) = f_1(g_1(x;T_{fm});T_{mf}).
\label{defsbif1}
\end{equation}
And one can define the corresponding generating functions $G_0$ and $G_1$
for the vertices representing the females.

Using these generating functions, we can now calculate generating functions
$H_0$ and $H_1$ for the sizes of outbreaks of the disease in terms either
of number of women or of number of men affected.  The calculation proceeds
exactly as in the unipartite case, and the resulting equations for $H_0$
and $H_1$ are identical to Eqs.~\eref{bondh1} and~\eref{bondh0}.  We can
also calculate the average outbreak size and the size of an epidemic
outbreak, if one is possible, from Eqs.~\eref{avs}, \eref{giant2},
and~\eref{giant3}.  The average outbreak size for males, for example, is
\begin{eqnarray}
\av{s} &=& 1 + {F_0'(1;T_{mf},T_{fm})\over
           1 - F_1'(1;T_{fm},T_{mf})}\nonumber\\
       &=& 1 + {T_{mf} T_{fm} f_0'(1) g_1'(1)\over
           1 - T_{mf} T_{fm} f_1'(1) g_1'(1)}.
\label{biavs}
\end{eqnarray}
The epidemic transition~takes place when $F_1'(1;T_{mf},T_{fm}) = 1$, or
equivalently when
\begin{equation}
T_{mf} T_{fm} f_1'(1) g_1'(1) = 1,
\end{equation}
and hence the epidemic threshold takes the form of a hyperbola in
$T_{mf}$--$T_{fm}$ space:
\begin{equation}
T_{mf} T_{fm} = {1\over f_1'(1) g_1'(1)}
              = {\mu\nu\over\sum_j j(j-1)p_j \sum_k k(k-1)p_k}.
\label{bitc}
\end{equation}
Note that this expression is symmetric in the variables describing the
properties of males and females.  Although we derived it by considering the
generating function for males $F_1$, we get the same threshold if we
consider $G_1$ instead.  Eq.~\eref{biavs} is not symmetric in this way, so
that the typical numbers of males and females affected by an outbreak may
be different.  On the other hand Eq.~\eref{bitc} involves the
transmissibilities $T_{mf}$ and $T_{fm}$ only in the form of their product,
and hence the quantities of interest are a function only of a single
variable $T_{mf}T_{fm}$.

The generalizations of Section~\ref{correlated}, where we considered
transmission probabilities that vary from one vertex to another, are
possible also for the bipartite graph considered here.  The derivations are
straightforward and we leave them as an exercise for the reader.

\begin{figure}
\resizebox{\figurewidth}{!}{\includegraphics{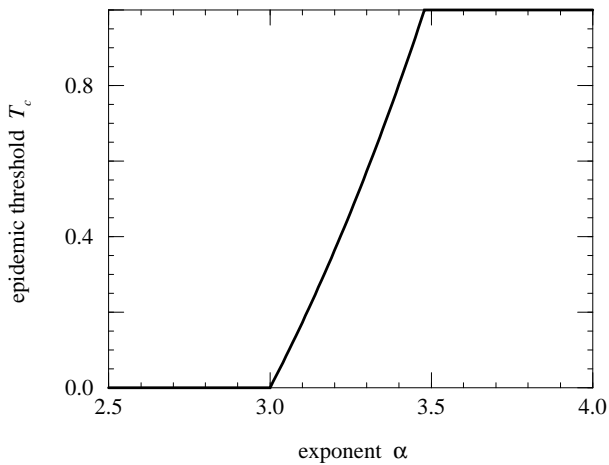}}
\caption{The critical transmissibility $T_c$ for the model of a sexually
  transmitted disease discussed in the text.  $T_c$ is greater than zero
  and less than one only in the small range $3<\alpha<3.4788$ of the
  exponent~$\alpha$.}
\label{tcfig}
\end{figure}

\subsection{Discussion}
One important result that follows immediately from Eq.~\eref{bitc} is that
if the degree distributions are truly power-law in form, then there exists
an epidemic transition only for a small range of values of the exponent of
the power law.  Let us assume, as appears to be the case, that the
exponents are roughly equal for men and women: $\alpha_m=\alpha_f=\alpha$.
Then Eq.~\eref{bitc} tells us that the epidemic falls on the hyperbola
$T_{mf} T_{fm} = T_c^2$, where
\begin{equation}
T_c = {\zeta(\alpha-1)\over\zeta(\alpha-2)-\zeta(\alpha-1)},
\end{equation}
where $\zeta(x)$ is the Riemann $\zeta$-function.  The behavior of $T_c$ as
a function of $\alpha$ is depicted in Fig.~\ref{tcfig}.  As the figure
shows, if $\alpha\le3$, $T_c=0$ and hence $T_{mf} T_{fm} = 0$, which is
only possible if at least one of the transmissibilities $T_{mf}$ and
$T_{fm}$ is zero.  As long as both are positive, we will always be in the
epidemic regime, and this would clearly be bad news.  No amount of
precautionary measures to reduce the probability of transmission would ever
eradicate the disease.  (Lloyd and May~\cite{LM01} have pointed out that a
related result appears in the theory of fully mixed models, where a
heterogeneous distribution of the infection parameter~$\beta$ (see
Eq.~\eref{sir}) with a divergent coefficient of variation will result in
the absence of an epidemic threshold.  Pastor-Satorras and
Vespignani~\cite{PV01a} have made similar predictions using mean-field-like
solutions for SIRS-type endemic disease models on networks with power-law
degree distributions and a similar result has also been reported for
percolation models by Cohen~\etal~\cite{CEBH00}.)  Conversely, if
$\alpha>\alpha_c$, where $\alpha_c=3.4788\ldots$ is the solution of
$\zeta(\alpha-2)=2\zeta(\alpha-1)$, we find that $T_c=1$ and hence $T_{mf}
T_{fm}=1$, which is only possible if both $T_{mf}$ and $T_{fm}$ are~1.
When either is less than~1 no epidemic will ever occur, which would be good
news.  Only in the small intermediate region $3<\alpha<3.4788$ does the
model possess an epidemic transition.  Interestingly, the real-world
network measured by Liljeros~\etal~\cite{Liljeros01} appears to fall
precisely in this region, with $\alpha\simeq3.2$.  If true, this would be
both good and bad news.  On the bad side, it means that epidemics can
occur.  But on the good side, it means that it is in theory possible to
prevent an epidemic by reducing the probability of transmission, which is
precisely what most health education campaigns attempt to do.  The
predicted critical value of the transmissibility is $T_c=0.363\ldots$ for
$\alpha=3.2$.  Epidemic behavior would cease were it possible to arrange
for the transmissibility to fall below this value.

Some caveats are in order here.  The error bars on the values of the
exponent $\alpha$ are quite large (about $\pm0.3$~\cite{Liljeros01}).
Thus, assuming that the conclusion of a power-law degree distribution is
correct in the first place, it is still possible that $\alpha<3$, putting
us in the regime where there is always epidemic behavior regardless of the
value of the transmissibility.  (The error bars are also large enough to
put us in the regime $\alpha>\alpha_c$ in which there are no epidemics.
Empirical evidence suggests that the real world is not in this regime
however, since epidemics plainly do occur.)

It is also quite possible that the distribution is not a perfect power law.
Although the measured distributions do appear to have power-law tails, it
seems likely that these tails are cut off at some point.  If this is the
case, then there will always be an epidemic transition at finite~$T$,
regardless of the value of~$\alpha$.  Furthermore, if it were possible to
reduce the number of partners that the most active members of the network
have, so that the cutoff moves lower, then the epidemic threshold rises,
making it easier to eradicate the disease.  Interestingly, the fraction of
individuals in the network whose degree need change in order to make a
significant difference is quite small.  At $\alpha=3$, for instance, a
change in the value~$\kappa$ of the cutoff from $\kappa=\infty$ to
$\kappa=100$ affects only 1.3\% of the population, but increases the
epidemic threshold from $T_c=0$ to $T_c=0.52$.  In other words, targeting
preventive efforts at changing the behavior of the most active members of
the network may be a much better way of limiting the spread of disease than
targeting everyone.  (This suggestion is certainly not new, but our models
provide a quantitative basis for assessing its efficacy.)

Another application of the techniques presented here is described in
Ref.~\onlinecite{ANMS02}.  In that paper we model in detail the spread of
walking pneumonia ({\em Mycoplasma pneumoniae\/}) in a closed setting (a
hospital) for which network data are available from observation of an
actual outbreak.  In this example, our exact solutions agree well both with
simulations and with data from the outbreak studied.  Furthermore,
examination of the analytic solution allows us to make specific suggestions
about possible new control strategies for {\em M.  pneumoniae\/} infections
in settings of this type.

\section{Conclusions}
\label{concs}
In this paper, we have shown that a large class of the so-called SIR models
of epidemic disease can be solved exactly on networks of various kinds
using a combination of mapping to percolation models and generating
function methods.  We have given solutions for simple unipartite graphs
with arbitrary degree distributions and heterogeneous and possibly
correlated infectiveness times and transmission probabilities.  We have
also given one example of a solution on a structured network---the spread
of a sexually transmitted disease on a bipartite graph of men and women.
Our methods provide analytic expressions for the sizes of both epidemic and
non-epidemic outbreaks and for the position of the epidemic threshold, as
well as network measures such as the mean degree of individuals affected in
an epidemic.

Applications of the techniques described here are possible for networks
specific to many settings, and hold promise for the better understanding of
the role that network structure plays in the spread of disease.

\begin{acknowledgments}
  The author thanks Lauren Ancel, L\'aszl\'o Barab\'asi, Duncan Callaway,
  Michelle Girvan, Catherine Macken, Jim Moody, Martina Morris, and Len
  Sander for useful comments.  This work was supported in part by the
  National Science Foundation under grant number DMS--0109086.
\end{acknowledgments}

\end{document}